\documentclass[journal=jacsat,manuscript=article]{achemso}

\usepackage{graphicx}
\usepackage{dcolumn}
\usepackage{bm}
\usepackage{multirow}
\usepackage{color}
\usepackage[normalem]{ulem}
\usepackage{amsmath}
\usepackage{gensymb}
\usepackage[colorlinks=true]{hyperref}
\usepackage{array}
\newcolumntype{P}[1]{>{\centering\arraybackslash}p{#1}}

\author{Meiling Zhang$^1$}
\affiliation{CEMES-CNRS, Universit\'e de Toulouse, CNRS, 31055 Toulouse, France}
\author{Jean-Marie Poumirol$^1$}
\affiliation{CEMES-CNRS, Universit\'e de Toulouse, CNRS, 31055 Toulouse, France}
\email{jean-marie.poumirol@cemes.fr}
\author{Nicolas Chery}
\affiliation{CEMES-CNRS, Universit\'e de Toulouse, CNRS, 31055 Toulouse, France}
\author{Cl\'ement Majorel}
\affiliation{CEMES-CNRS, Universit\'e de Toulouse, CNRS, 31055 Toulouse, France}
\author{R\'emi Demoulin}
\affiliation{Normandie Univ, UNIROUEN, INSA Rouen, CNRS, Groupe de Physique des Mat\'eriaux, 76000 Rouen, France}
\author{Etienne Talbot}
\affiliation{Normandie Univ, UNIROUEN, INSA Rouen, CNRS, Groupe de Physique des Mat\'eriaux, 76000 Rouen, France}
\author{Hervé Rinnert}
\affiliation{Universit\'e de Lorraine CNRS, IJL, Nancy, France}
\author{Christian Girard}
\affiliation{CEMES-CNRS, Universit\'e de Toulouse, CNRS, 31055 Toulouse, France}
\author{Filadelfo Cristiano}
\affiliation{LAAS-CNRS, Universit\'e de Toulouse, CNRS,  31031 Toulouse, France}
\author{Peter R. Wiecha}
\affiliation{LAAS-CNRS, Universit\'e de Toulouse, CNRS,  31031 Toulouse, France}
\author{Teresa Hungria}
\affiliation{Centre de Microcaract\'erisation Raimond Castaing (UAR 3623), 31400 Toulouse, France}
\author{Vincent Paillard}
\affiliation{CEMES-CNRS, Universit\'e de Toulouse, CNRS, 31055 Toulouse, France}
\author{Arnaud Arbouet}
\affiliation{CEMES-CNRS, Universit\'e de Toulouse, CNRS, 31055 Toulouse, France}
\author{Béatrice P\'ecassou}
\affiliation{CEMES-CNRS, Universit\'e de Toulouse, CNRS, 31055 Toulouse, France}
\author{Fabrice Gourbilleau}
\affiliation{CIMAP, Normandie Univ, ENSICAEN, UNICAEN, CEA, CNRS, 6 Boulevard Mar\'echal Juin, 14050, Caen Cedex 4, France}
\author{Caroline Bonafos}
\email{caroline.bonafos@cemes.fr}
\affiliation{CEMES-CNRS, Universit\'e de Toulouse, CNRS, 31055 Toulouse, France}

\title{Infrared nanoplasmonic properties of hyperdoped embedded Si nanocrystals in the few electrons regime}


\begin{document}
\footnotemark{contributed equally to this paper}
\begin{abstract}

Using Localized Surface Plasmon Resonance (LSPR) as an optical probe we demonstrate the presence of free carriers in phosphorus doped silicon nanocrystals (SiNCs) embedded in a silica matrix. In small SiNCs, with radius ranging from 2.6 to 5.5 nm, the infrared spectroscopy study coupled to numerical simulations allows us to determine the number of electrically active phosphorus atoms with a precision of a few atoms. We demonstrate that LSP resonances can be supported with only about 10 free electrons per nanocrystal, confirming theoretical predictions and probing the limit of the collective nature of plasmons. We reveal a phenomenon, unique to embedded nanocrystals, with the appearance of an avoided crossing behavior linked to the hybridization between the localized surface plasmon in the doped nanocrystals and the silica matrix phonon modes.  Finally, a careful analysis of the scattering time dependence versus carrier density in the small size regime allows us to detect the appearance of a new scattering process at high dopant concentration. 

 \end{abstract}

\maketitle

Localized surface plasmon resonances (LSPR), the collective oscillations of free carriers at metal surfaces, have been studied for decades due to their ability to fundamentally alter light-matter interactions and the resulting remarkable applications in: enhanced spectroscopy \cite{kneipp_single_1997}, sensing \cite{larsson_elin_m_nanoplasmonic_2009}, optical devices, and photovoltaics \cite{atwater_plasmonics_2010}. For a long time, plasmonics-based technology relied heavily on noble metal nanostructures due to their intense LSPR in the visible range \cite{link_size_1999}. Things have begun to change recently, with the observation of LSPR in doped semiconductors. The new form of control provided by the tunability of carrier density in semiconductors, combined with the resulting expansion of plasmonics into the Infrared spectral range (not easily reached with metals), has fueled the search for novel plasmonic materials with improved crystalline quality, integrability, tunability, and potentially lower losses \cite{naik_alternative_2013, poumirol_hyper-doped_2021}.
 
However the real paradigm shift came with the observation of LSPR in very small semiconducting nanostructures and nanocrystals \cite{zhou_comparative_2015, luther_localized_2011, hunter_probing_2020,kramer_plasmonic_2015,rowe_phosphorus-doped_2013}. Because the LSPR frequency becomes more sensitive to changes in the number of free carriers as the size of the nanocrystal decreases, with electro-optic sensitivity going as $1/r^{3}$ (r being the radius of the NC), it is possible to detect the addition of a very small number of carriers to a nanocrystal. In this context, LSPRs have emerged as an optical probe (contactless) with unprecedented sensitivity to processes involving carrier dynamics, that would otherwise be difficult to access. Furthermore, the concept of plasmon supported by a small number of carriers raises fundamental questions that go beyond material science issues, such as the minimum number of charge carriers required to sustain a plasmon resonance, the effect of confinement on the plasmon resonance, and the validity of the Drude model in such conditions.

In this article we investigate the plasmonic properties of phosphorus-doped Si nanocrystals (SiNCs) embedded in SiO$_2$. We demonstrate that SiNCs can be massively doped by combining Low Energy Ion Implantation and Rapid Thermal Annealing (RTA), with measured LSPR wavelength ranging from 4 to 7 \,$\mu$m depending on the P implantation dose. To determine the carrier density and electronic scattering time for all experimental doses, we use  homemade numerical simulations based on the Green Dyadic Method (GDM), taking into account the complex multilayered dielectric environment of our samples. The simulation allowed us to put into light an hybridization between the LSPR and the optical phonon modes of the silica matrix. We were able to determine the variation of the phosphorus activation rate and pinpoint a lower limit to the number of activated carriers required to support plasmonic resonances by combining these results with atomically resolved 3D mapping. Finally, we demonstrate three distinct phenomena that drive the electronic scattering processes in SiNCs.

\begin{figure}
	\includegraphics*[width=10.5cm]{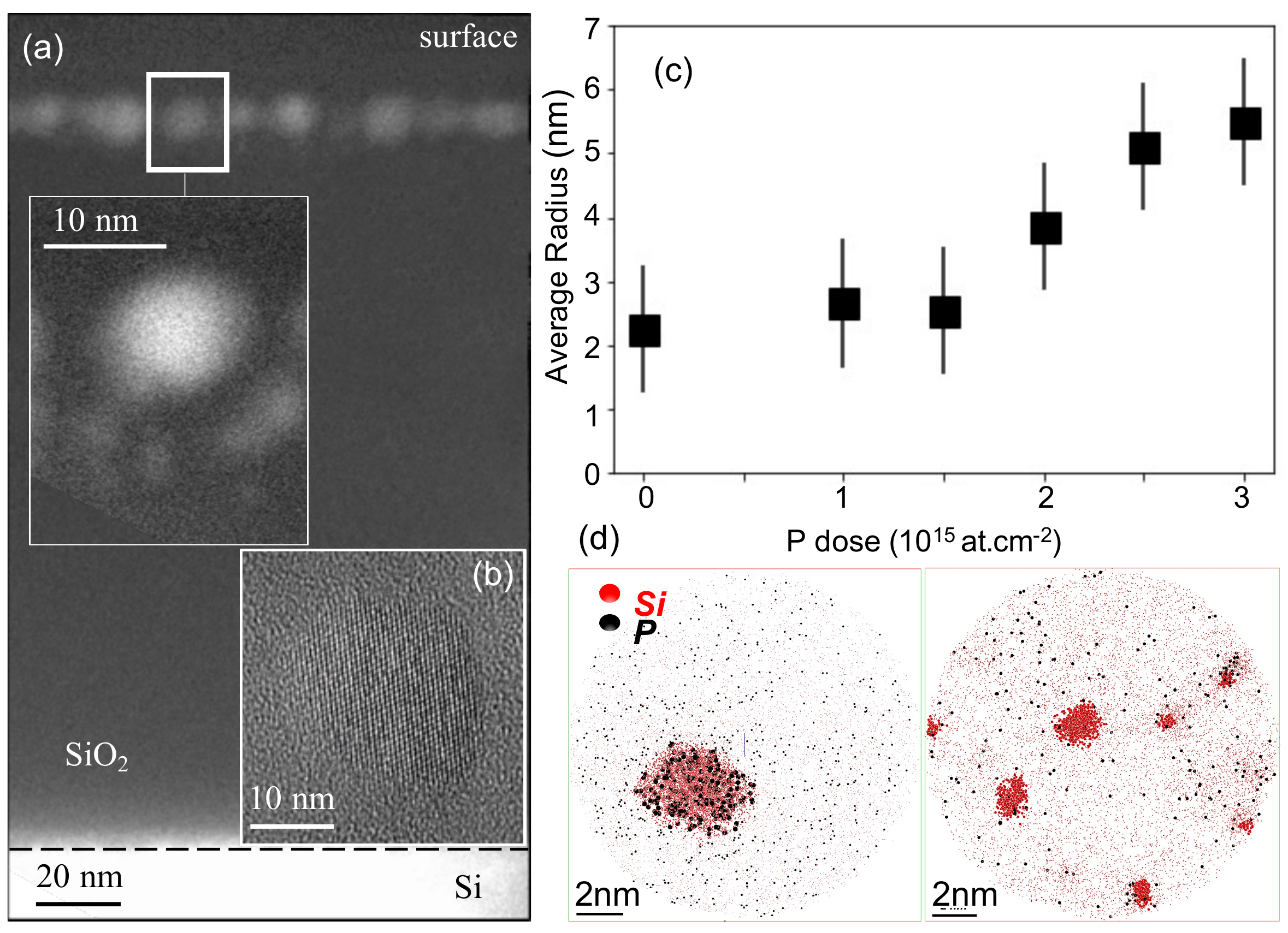} 
	\caption{(a) EFTEM cross-sectional image displaying the spatial distribution of doped SiNCs for sample P5 (implanted at 3 $\times$ 10$^{15}$ at.cm$^{-3}$), Inset: Zoom on the 20nm implantation layer. (b) HREM image of a SiNC in the (110) zone axis. (c) Evolution of the size (big SiNCs) with implanted dose. (d) APT images of a large NC located at the projected range of the Si implantation profile (left panel) and of small NCs located in the profile tail (right panel). The red (black) dots are the Si (P) atoms respectively.} 
	\label{fig1}
\end{figure}


\section{Results.}
\subsection{SiNCs fabrication and doping.}
A 200 nm thick thermal silica layer on top of a silicon substrate is implanted with Si ions at an energy of 8 keV and a dose of 2$\times$10$^{16}$ cm$^{-2}$. An intermediate RTA is performed for 60 seconds at 1050$^{o}$C in N$_2$ with the goal of phase separation and SiNCs formation. After the synthesis of SiNCs, dopant impurities are introduced into the SiNCs to decouple equilibrium properties from kinetic effects \cite{arduca_doping_2017}. Phosphorus is implanted in 5 doses at 7 keV to match the projected Si range. The ratio of the P and Si doses ranges from 5 to 15 $\%$. Following P implantation, the samples undergo a new annealing at 1050$^{o}$C for 60 seconds dedicated to dopant activation. As a last step, a forming gas (FG) annealing at N$_2$+10 $\%$ H$_2$ was used to passivate dangling bonds at the SiNCs/matrix interface \cite{fernandez2002}.

Figure \ref{fig1}(a) shows a low magnification Energy Filtered TEM (EFTEM) image of the SiNCs after doping. They are distributed along a 20 nm-thick band located at 4 nm from the surface. The doped SiNCs show a sharp and bimodal size distribution (see SI Fig. S1), with the largest SiNCs located at the profile projected range (15 nm) and small NCs located deeper in the profile tail. Figure \ref{fig1}(b) a high resolution image clearly shows that the doped SiNCs are spherical and remain crystalline after doping, with the same lattice parameter as bulk Si. At doses up to 1.5$\times$10$^{15}$ cm$^{-2}$, the average NC diameter remains constant for both small (2.8 nm) NCs located in the profile tail and large (5 nm) NCs formed at the maximum of the implantation profile. On the contrary, at higher doses, the average diameter of the big NCs increases with the P dose, reaching up to twice the size of the low dose regime (see Fig. \ref{fig1}(c)). The increase in P concentration has no effect on the size of the small SiNCs. This enhanced NC growth with P concentration has previously been observed in the literature for SiNCs in a silica matrix prepared using various techniques \cite{hao_effects_2009, yang_properties_2017,fujii_below_2003}. The hypothesis of P doping softening silica matrices has been proposed, resulting in longer diffusion length of Si atoms during annealing, and larger particles \cite{fujii_below_2003}. Enhanced diffusion by implanted P atoms has also been proposed as a possible explanation for the favored SiNCs phase separation in the presence of P \cite{hao_effects_2009}. A careful examination of atomically resolved 3D images obtained by laser-assisted Atom Probe Tomogrphy (ATP), shown in Figure \ref{fig1}(d) reveals the precise location of the dopant. The majority of P atoms, black dots on the figure, are well-concentrated inside the big SiNCs (see left panel), with a homogeneous distribution and no accumulation of dopant at the surface. These findings, are consistent with theoretical calculations \cite{guerra_preferential_2014, garcia-castello_energetics_2015-1} and previous observations for similar systems \cite{khelifi_efficient_2013, demoulin_atomic-scale_2019, nomoto_atom_2016}. The position of P in the core of the SiNCs contradicts self-purification theories but is consistent with the well-known macroscopic behavior of these dopants at bulk Si/SiO$_2$ interfaces, where segregation is in favor of silicon for phosphorus at thermal equilibrium \cite{sakamoto_segregation_1987}. Small NCs contain very few, if no, P atoms, which is not surprising as they are located deeper inside the SiO$_{2}$ matrix in the profile tail of the P implantation profile (see right panel Fig. \ref{fig1}(d)). 

Scanning Transmission Electron Microscopy-Energy Dispersive X-Ray Spectroscopy (STEM-EDX) was used to determine the P content of SiNCs. The phosphorus Cliff-Lorimer coefficient has been quantitatively calibrated on a reference sample (P implantation in bulk Si) \cite{chery_study_2022} . The P concentration in big SiNCs ranges from 3.2 to 6.3 $\times$ 10$^{21}$ cm$^{-3}$ (see Fig. \ref{fig:5}(b)), whereas it is negligible and close to the detection limit in the small SiNCs and in the matrix. The high P concentration of the large SiNCs is consistent with previous studies \cite{khelifi_efficient_2013, perego_thermodynamic_2015,demoulin_atomic-scale_2019} that conclude that doping SiNCs in SiO$_2$ corresponds to a thermodynamically advantageous configuration. The measured P concentrations are up to 6 times higher than the solid solubility of P in bulk silicon at 1050$^{o}$C (10$^{21}$ cm$^{-3}$) \cite{olesinski_psi_1985}, implying that solubility in Si nanostructures has increased. \cite{perego_thermodynamic_2015, demoulin_atomic-scale_2019}

\begin{figure}[]
	\includegraphics*[width=8.5cm]{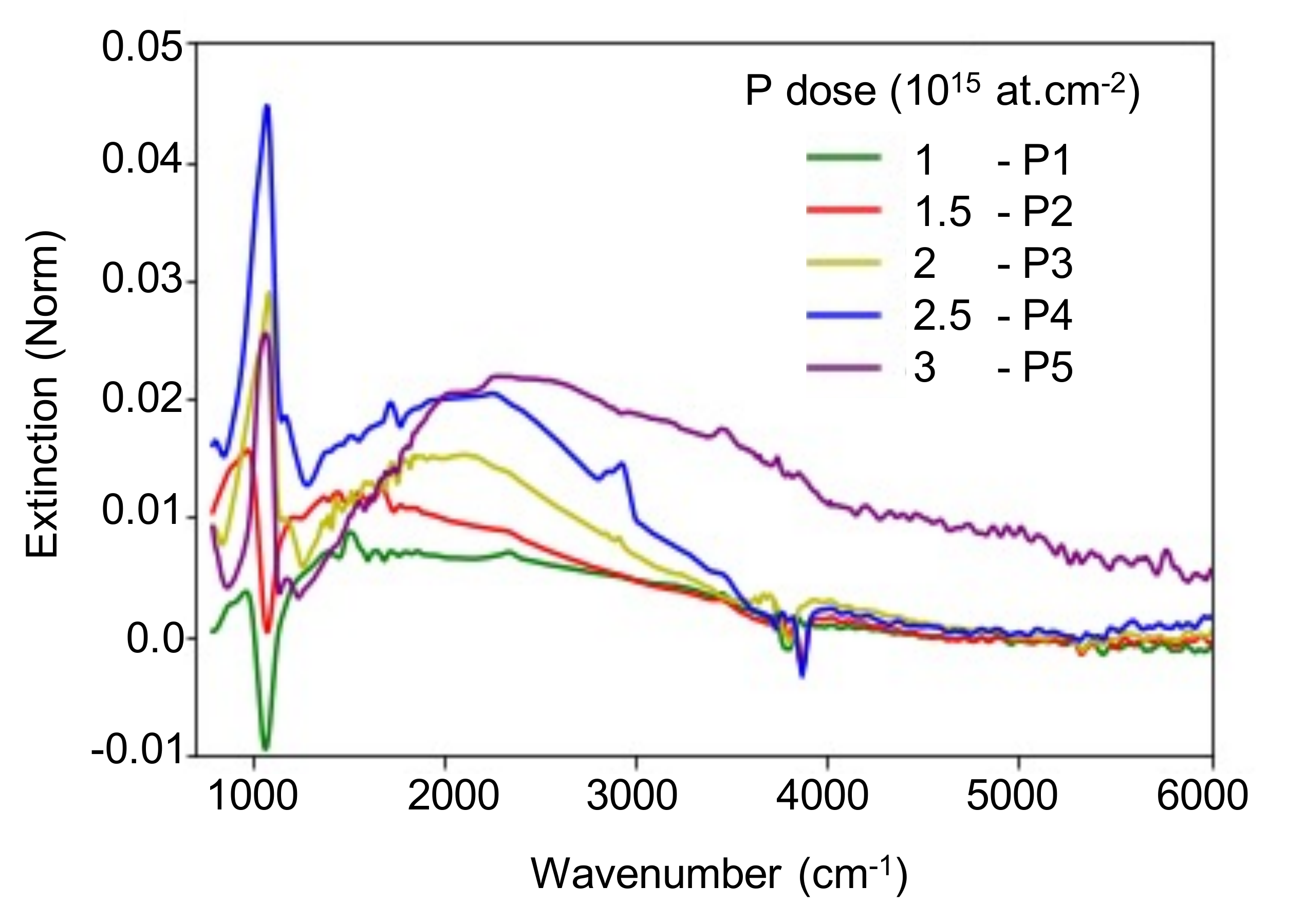} 
	\caption{Normalized Extinction curves measured for five different P doses.} 
	\label{fig:3}
\end{figure}


\subsection{SiNCs Plasmonic properties.} 
To determine the effect of the implanted phosphorus on the electronic properties of the SiNCs we performed Fourier Transform Infrared (FTIR) Spectroscopy at room temperature, in vacuum. This technique is an all-optical, non destructive and very sensitive method for active dopant concentration measurement in semiconductor bulk and nanostructures \cite{poumirol_hyper-doped_2021}. Figure \ref{fig:3} shows the normalized extinction spectra recorded for the different implanted doses (P1 to P5).  T. The extinction spectra are normalized using undoped SiNCs (P0) sample as a reference. This normalization procedure reduces the optical signature of the thermal silica layer as well as the likelihood of seeing interference patterns in the spectra.  Two interesting features can be noticed in all five extinction curves: (i) A relatively sharp feature that appears at low frequency (around 1050 cm$^{-1}$) and can be attributed to the multiple SiO$_2$ phonon modes that are not completely removed by normalization \cite{malitson_interspecimen_1965,kitamura_optical_2007}. (ii) A broad peak that shifts progressively higher in frequency as the P dose, thus the free carrier density, increases. This maximum of extinction is most likely related to the excitation of a localized surface plasmon in doped SiNCs. The amplitude of the plasmon mode is quite small, as expected for such small NC, reaching 2$\%$ for the highest dose (P5) and falling to 0.5$\%$ for the lowest dose (P1). Nonetheless, it is plainly visible. 

\begin{figure}[]
	\includegraphics*[width=10.5cm]{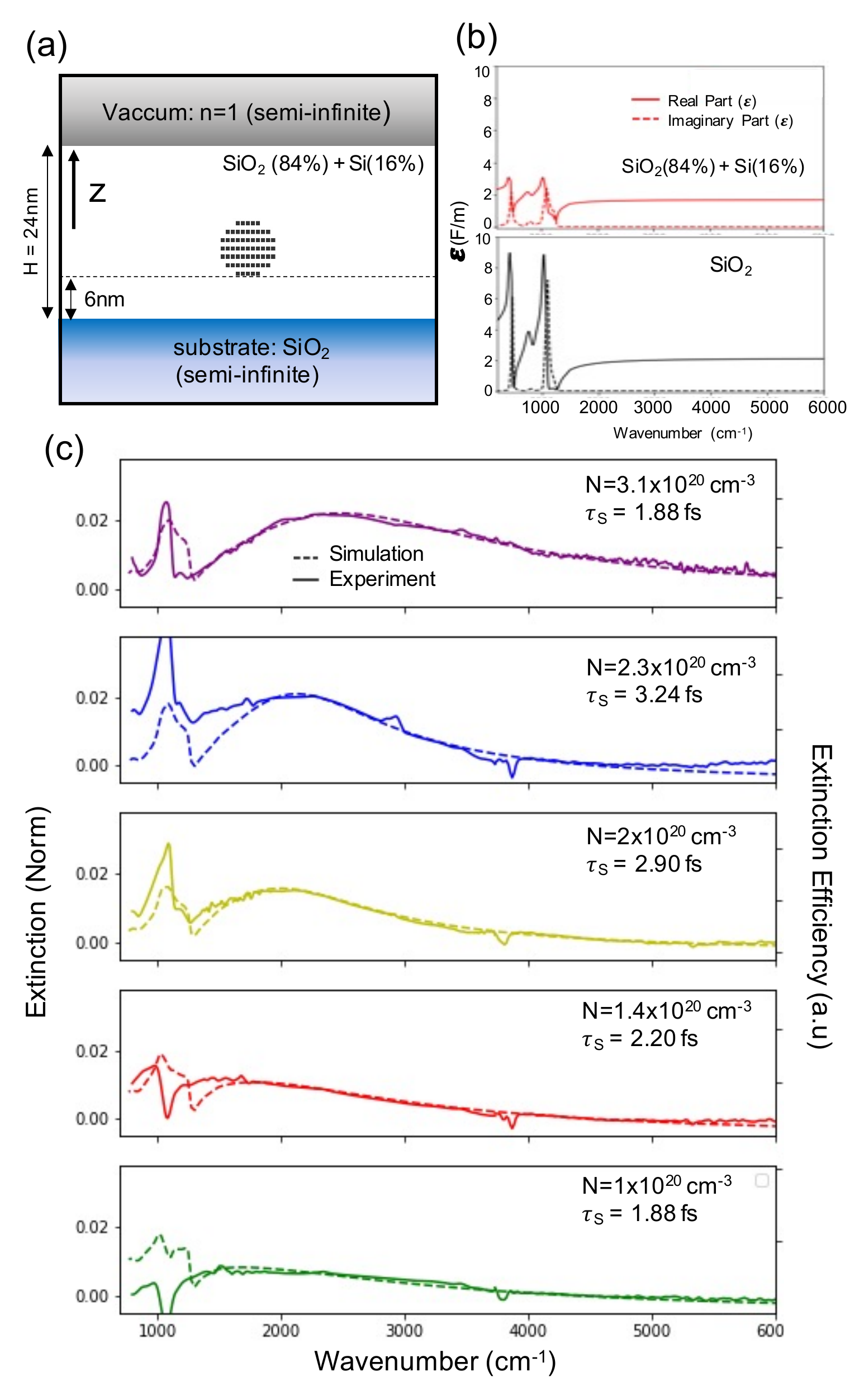} 
	\caption{(a) Geometry used in the numerical simulations. the light is propagating along the -z direction. (b) Complex dielectric functions used in calculations. Top: calculated from equation \ref{eq3}. Bottom: pure silica from \cite{malitson_interspecimen_1965,kitamura_optical_2007}. (c) Comparison between simulated normalized extinction (dashed line) and experimental measurement (continuous line). The resulting fitting parameters are shown in the top right corner for each P dose.} 
	\label{fig:4}
\end{figure}

To confirm the plasmonic nature of the observed modification of extinction we model our experimental results using numerical simulations based on the Green Dyadic Method (GDM) with our homemade simulation toolkit “pyGDM” \cite{wiecha_pygdmpython_2018, majorelTheoryPlasmonicProperties2019}. This method, which is based on nanostructure volume discretization, allows us to simulate the complex multilayered geometry of our samples, as shown in Figure \ref{fig1}(a). Figure \ref{fig:4} (a) depicts a schematic of the simulated geometry, with the NC represented by a doped silicon sphere. The Drude-Lorentz model is used to describe the dielectric function of the doped Si layer:
%
\begin{equation}
\epsilon_{\text{Si}}^{doped} (\omega) = \epsilon_{\text{Si}}^{undoped}(\omega) - \frac{\omega^2_{\text{p}}}{\omega(\omega+i/\tau)} \, ,
\label{eq1}
\end{equation}

where $\omega_{\text{p}}$ = $\sqrt{4\pi Ne^2/m^{*}}$ the Drude weight, $\tau$ is the electronic scattering rate and $\epsilon_{\text{Si}}^{undoped}(\omega)$ the dielectric function of intrinsic Si taken from Ref \cite{chandler-horowitz_high-accuracy_2005}. In this expression, $m^*= 0.3m_e$ is the electron effective mass in Si, $e$ the elementary charge, $N$ the free carrier concentration. For each P dose the radius of the sphere is chosen to be the average radius of the large SiNCs (Figure \ref{fig1}(c)) in the bimodal size-distribution (see supplementary information, Figure S1). The doped Si sphere is then placed inside a 24 nm thick silicon rich SiO$_2$ layer. The small NCs contained in this layer are undoped and thus do not contribute directly to the observed plasmonic properties, nevertheless they are made of pure silicon and present a dielectric function drastically different from the rest of the silica matrix. To accurately describe the complex dielectric environment of this layer we use the Maxwell Garnett mixing formula. It has been developed to approximate a complex electromagnetic medium such as a colloidal solution of gold micro-particles and gives an effective permittivity in terms of the permittivities and volume fractions of the individual constituents of the complex medium \cite{markel_introduction_2016, garnett_xii_1904}: 

\begin{equation}
\epsilon_{layer} (\omega)= \epsilon_{SiO_2} \frac{\epsilon_{SiO_2}+\frac{1+2f}{3}(\epsilon_{Si}^{undoped}-\epsilon_{SiO_2})}{\epsilon_{SiO_2}+\frac{1-f}{3}(\epsilon_{Si}^{undoped}-\epsilon_{SiO_2})}
\label{eq3}
\end{equation}
With $\emph{f}$ the volume fraction of Si inclusions. In our calculation we fix $\emph{f}$ to 16$\%$, based on STEM-EDX measurement, the resulting dielectric function is displayed in Figure \ref{fig:4}(b) (top panel). Finally this layer is sandwiched between a SiO$_{2}$ substrate (see dielectric function in Figure \ref{fig:4}(b) (bottom panel) and a vacuum cladding. The extinction efficiencies of individual doped SiNC in the simulated environment are then calculated using the Lippmann-Schwinger equation. 

\begin{figure}[]
	\includegraphics*[width=8.5cm]{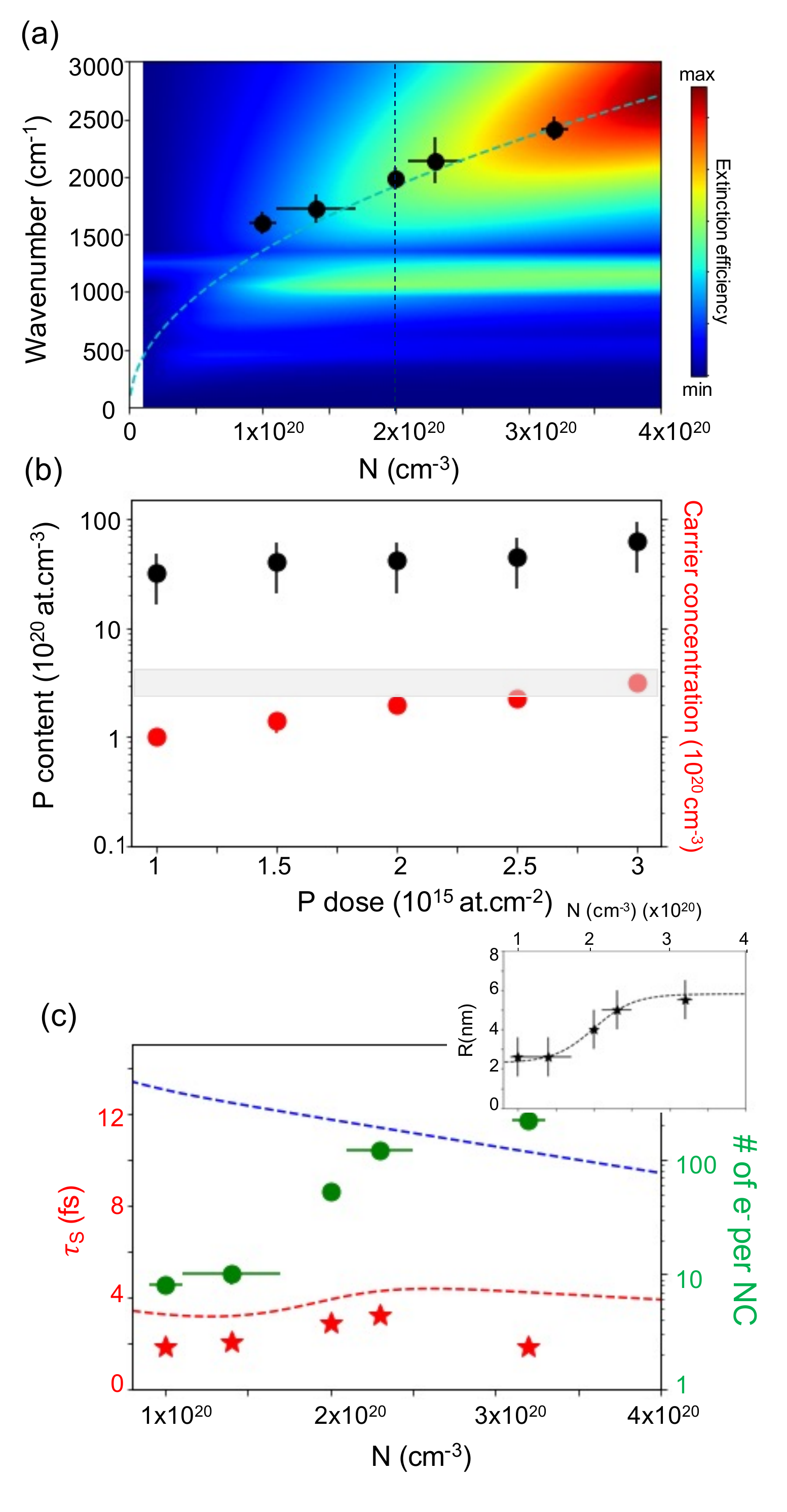} 
	\caption{(a) Energy of the LSPR ($\omega_P$) versus carrier density (black dots) deduced from the fitting procedure depicted in Figure \ref{fig:4}. Color map: GDM calculations, frequency dependent extinction versus carrier density. Dashed blue line: exact Mie solution for spherical particle (from equation \ref{eq2}). Vertical dashed line: Cut of the color map along this line is the yellow curve shown in Fig. \ref{fig:4} (c). (b) P concentration inside big particles measured by STEM-EDX (black dots) and carrier concentration measured by FTIR (red circles) as a function of P dose. The range of experimental values for electrical solubility of P in bulk Si at 1050$^{o}$C is represented by the grey rectangle \cite{pichler_intrinsic_2004}. (c) Left axis: scattering time versus carrier density extracted from fitting procedure (red stars). Dashed blue line: $\tau_{bulk}(N)$\cite{masetti_modeling_1983}. Red dashed line: $\tau_{eff}(N)$ deduced from equation \ref{eq5}. Right axis: average number of electrons  in each NC contributing to the plasmon resonance versus carrier density (green circles). Inset: Black dot: Measured average size of the doped Si-NCs versus carrier density. Dashed black line: $r(N)$ used in equation \ref{eq2}.} 
	\label{fig:5}
\end{figure}
Figure \ref{fig:4}(c) depicts the best fit we obtained for all P implanted doses. Note that as we calculate the extinction of one single NC, the simulation is not expected to reproduce the extinction absolute amplitude. All simulated curves in Figure \ref{fig:4}(c) are by consequence normalized to the experimental LSPR amplitude. The fitting parameters are the carrier density $N$ and scattering time $\tau$, with all other parameters held constant. For all samples, the simulated extinctions reproduce the LSPR signature very well. The theoretical description of the phonon-related extinction is satisfactory but not perfect. This result is not entirely unexpected given that the Maxwell Garnett model we used to simulate the Si rich silica layer (see equation \ref{eq3}) assumes isotropic composites i.e. with equivalent nanoparticle densities in all directions in space equivalent. However the undoped SiNCs inside the silica matrix are not uniformly distributed around the doped NC (see inset Figure \ref{fig1} (a)). Nevertheless, the quality of the fitting procedure is a strong indication that the approximation is quite good (as a reference Figure S3 in supplementary shows the same simulations using $\epsilon_{layer} = \epsilon_{SiO_2}$). 

Figure \ref{fig:5} (a) depicts the maximum of extinction versus $N$ (black dots) extracted from Figure \ref{fig:4}(c). As a reference, we plot in dashed line, on the same graph, the expected LSPR energy $\omega_{sp}$ (Mie resonance) for a spherical NC  \cite{maier_plasmonics_2007}:
\begin{equation}
\hbar \omega_{sp} = \sqrt{\frac{N_{\text{e}}e^2}{\epsilon_0 m^{*}( \epsilon_\infty^{\text{Si}}+2\epsilon_{\text{m}})}} \, ,
\label{eq2}
\end{equation}
where $\epsilon_{\text{m}}$ , the average dielectric function of the surrounding environment, is taken to be $\epsilon_{layer}(\omega=6000cm^{-1})$ = 2.3 far from any phonon mode. One can see that at high carrier density the experimental points follow the expected square root dependence for plasmon resonances, but start deviating from it at low carrier density. To understand this discrepancy we calculated the frequency dependent extinction versus carrier density with a constant sphere radius of $r = 4$ nm and constant scattering time $\tau =2.9$ fs, shown as the superimposed color map on the figure. This simulation indicates clearly that this deviation is mainly due to a hybridization between phonons of silica and the LSPR resulting in an avoided crossing \cite{marty_plasphonics_2013, abstreiter_light_1984}. The very good description of the experimental results we obtained here, while using a single value of NC's radius ($r = 4$ nm) for all carrier densities, discard possible explanations linked to the NC's size, in agreement with theoretical prediction that estimate that no quantization effects should be observed for Si-NCs with $r \ge 2nm$ \cite{pi_tightbinding_2013}. Using the carrier density and average volume of the NC in our samples, we can calculate the average number of electrons (equal to the number of activated P atoms) in each NC contributing to the plasmon resonances. We deduced these values with a precision of few electrons and they range from more than 100 electrons for larger NCs (higher P dose) down to approximately ten electrons for small NCs (lower P dose, see green circles in Fig. \ref{fig:5}(c)). Due to the collective nature of the plasmonic resonance the question of the minimum number of required electrons is of fundamental nature, our work is confirming the theoretical prediction made in \cite{pi_tightbinding_2013} that estimated a minimum of 10 electrons for the plasmonic mode to emerge. This value is quite similar to what has been observed in metallic clusters (Hg$_6^+$) which display plasmon-like features with 11 electrons. \cite{haberland_transition_1992}
To go further and evaluate the potential of these systems as IR plasmonic materials, we provide in SI a detailed evaluation of the achievable GPI in these systems (see Figure S5). \cite{zhang_how_2017}
 
It is worth noting  here that the degree of optical field confinement $\lambda_{IR}/(2nd)$ (where $d$ is the SiNC diameter and $n$ the optical index of $SiO_2$) observed in those samples is extremely high, ranging from 250 for P5 up to 800 for P1. This is significantly higher than what can be obtained with noble metals and even better than what has previously been observed in hyper doped Si based plasmonic metasurfaces \cite{poumirol_hyper-doped_2021}. 

In Figure \ref{fig:5}(b) the carrier density $N$ (red circles) is plotted as a function of the P implanted dose. One can clearly see that this value reaches the electrical solubility of P in bulk Si at $1050^o$C (grey rectangle) \cite{pichler_intrinsic_2004}. Beyond this value, the additional P atoms likely form small electrically inactive clusters. The activation rate, defined as the ratio of the carrier density $N$ to the total number of P atoms inserted inside the SiNCs measured in STEM-EDX ranges between 3 and 5 $\%$. This activation ratio is more than one to two orders of magnitude higher than that previously measured on freestanding \cite{gresback_controlled_2014} or silica embedded  \cite{gutsch_electronic_2015} P doped SiNCs of the same size. However, this activation efficiency is lower than that measured on Si nanostructures obtained through top down processing of hyperdoped Si overlayers, where the activation process occurs at the bulk state and is optimized by non-equilibrium melt laser annealing \cite{poumirol_hyper-doped_2021}.  



Figure \ref{fig:5} (c) shows the scattering time versus the extracted carrier density. For similar carrier density, the scattering time observed in SiNCs (red stars in Figure \ref{fig:5}(c)) is smaller than that observed in continuous Si layers (blue dashed line extracted from \cite{masetti_modeling_1983}). Furthermore, $\tau$(N) exhibits a non-monotonic dependence, increasing with $N$ at low density and decreasing at high density. To explain both observations, we must consider the effects of NC size on scattering time. As the diameter of the NC becomes smaller than the electron mean free path, electrons collide with the spherical NC surface at an average rate of  $2v_{F}/r$, where $v_{F}$ is the Fermi velocity in Si \cite{flytzanis_v_1991}. To account for the observed dependence of $\tau_{eff}(N)$, we must consider two competing effects:  (i) In silicon (bulk), the scattering time decreases with increasing carrier density, as described in Ref\cite{masetti_modeling_1983} (see $\tau_{bulk}(N)$ dashed blue line on figure \ref{fig:5}(c)), (ii) the size of the NC in our system is a function of the P content and therefore of the carrier density, with $r(N)$ an increasing function displayed in inset of Figure \ref{fig:5} (c). As a result, the effective collision time is N dependent \cite{commentaire}: 

\begin{equation}
1/\tau_{eff}(N)= 1/\tau_{bulk}(N)+ v_{F}(N)/r(N)
\label{eq5}
\end{equation}

with $v_{F}=\hbar/m^*(3\pi^2N)^{2/3} $. The resulting $\tau_{eff}(N)$ is plotted as a red dashed line in Figure \ref{fig:5}(c), and it accurately describes both the absolute values and the variation of the measured scattering time without the use of any adjusting parameters. It is now interesting to note that $\tau_{P5}$ exhibits a larger deviation from the predicted behavior. For this sample the inactive dopant concentration (difference between the total P concentration measured by STEM-EDX and the carrier density measured by FTIR), is multiplied by 1.5 (see supplementary Fig. S4). Hence, more and more of the implanted phosphorus begins aggregating and forms electrically inactive clusters of a few atoms, degrading the electronic scattering rate inside the SiNCs.  

In conclusion, we achieved massively doped SiNCs embedded in a silica matrix by combining Low Energy Ion Implantation and Rapid Thermal Annealing. We demonstrated that such embedded nanostructures can support localized surface plasmon resonances that are controlled by the silicon free carrier density. Despite a significant mismatch between the wavelength (4-7\,$\mu$m) and the Si-NC diameter ($\approx 4\,$nm), the LSPR allows a strong interaction with IR light, resulting in an increase in extinction of up to 2$\%$ at the resonance. We investigated the optical properties of a single nanocrystal in a complex dielectric environment using numerical simulations. We observed a deviation from the expected square root dependance of the LSPR with the carrier density that we attributed to a coupling between the plasmon mode and the infrared phonons of the surrounding silica matrix. We demonstrated that $\approx$ 10 free electrons inside NC are sufficient to produce LSPR. Finally, because the diameter of our NC is smaller than the electron mean free path, we show that the scattering time dependence with carrier density in such NCs differs significantly from what is observed in bulk or larger Si nanostructures, and we were able to detect evidence of P aggregate formation.

\section{Methods}
\subsubsection{Thermal Annealing} Rapid Thermal Annealing was carried out in a nitrogen atmosphere at a temperature of 1050$^o$ for 60 seconds. Halogen lamps allow to reach the targeted temperature with a speed of 20 $^o$C/s for up, and $\approx$10 $^o$C/s for down ramps. Forming gas annealing has been done using a H$_2$/N$_2$ mixture. This annealing process is dedicated to the passivation of the interface states between SiNCs and silica matrix . In this work, forming gas annealing using an RTA furnace was performed on pieces of samples at 500 $^o$ for 900 seconds with a gas mixture of 10$\%$ H$_2$ + 90$\%$ N$_2$.
\subsubsection{Simulations}
The Green Dyadic Method (GDM) is based on a volume discretization of the nanostructure and numerical solving of Maxwell's equations in the frequency domain. From the Lippmann-Schwinger equation:
\begin{equation}
\begin{split}  
\mathbf{E}(\mathbf{r},\omega) = \mathbf{E_0}(\mathbf{r},\omega) +
\frac{1}{4\pi} \int_{V_{\text{j}}}\big(\epsilon_{\text{Si}}(\omega) - 1) \mathbf{S} (\mathbf{r}, \mathbf{r}', \omega). \mathbf{E}(\mathbf{r'},\omega) d\mathbf{r}'
\end{split} 
\label{equation3}
\end{equation}
the time Fourier transform of the local electric field $\mathbf{E}(\mathbf{r},\omega)$ is self-consistently obtained at any location, as a function of the incident electric field $\mathbf{E}_0(\mathbf{r}',\omega)$. The field propagator $\mathbf{S} (\mathbf{r}, \mathbf{r}',\omega)$ allows to compute the extinction and scattering efficiencies as well as the near- and far-field radiation patterns of an individual doped Si-ND.\cite{girard_shaping_2008} 
The discretization step is small in order to accurately describe the spherical shape and optical properties of the NC. For the small SiNC observed at low P concentration (r=2.6 nm), we use a minimum of 150 dipoles on a hexagonal mesh and up to 600 dipoles for the high doses (r=5.5 nm) (see black squares in Fig. \ref{fig:4} (a)).

\bibliography{meiliing.bib}

\section{Acknowledgements} This work was partly funded by ANR DONNA (ANR-18-CE09-0034) and from the International Emerging Action from CNRS-DONNA . The authors thank the China Scholarship Council (201801810118) and l'Institut National de Physique of CNRS for funding. We thank Jérémie Teyssier and Dirk Van der Marel from UniGe for providing the IR spectrometer. The authors acknowledge the CALMIP computing facility (grant P12167) and Jean-Christophe Marrot and Eric Imbernon from the LAAS-CNRS micro and nanotechnologies platform, a member of the French RENATECH network for phosphorus implantations.

\section{Author contributions}

C.B, and F.G  planned the experiments. M.Z, and J.-M.P. carried out measurements. P.W, A.A and J.-M.P performed simulations. All authors discussed the data and contributed to writing the paper.

\section{Data availability}
All relevant data are available from the corresponding author upon request.

\end{document}